# RESEARCH ARTICLE

# Analysis of the Optical Properties and Electronic Structure of Semiconductors of the Cu₂NiXS₄ (X = Si, Ge, Sn) Family as New Promising Materials for Optoelectronic Devices

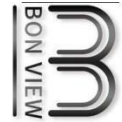

BON VIEW PUBLISHING

---

**Dilshod Nematov** [1,*]

[1] S.U. Umarov Physical-Technical Institute of the National Academy of Sciences of Tajikistan, Tajikistan

---

**Abstract:** In this work, the optoelectronic characteristics of kesterites of the Cu₂NiXS₄ system (X = Si, Ge, Sn) were studied. The electronic properties of the Cu₂NiXS₄ (X = Si, Ge, Sn) system were studied using first-principles calculations within the framework of density functional theory. For calculations, ab initio codes VASP and Wien2k were used. The high-precision modified Beke-Jones (mBJ) functional and the hybrid HSE06 functional were used to estimate the bandgap, electronic and optical properties. Calculations have shown that when replacing Si with Ge and Sn, the band gap decreases from 2.58 eV to 1.33 eV. Replacing Si with Ge and Sn reduces the overall density of electronic states. In addition, new deep (shallow) states are formed in the band gap of these crystals, which is confirmed by the behavior of their optical properties. The obtained band gap values are compared with existing experimental measurements, demonstrating good agreement between HSE06 calculations and experimental data. The nature of changes in the dielectric constant, absorption capacity and optical conductivity of these systems depending on the photon energy has also been studied. The statistical dielectric constant and refractive index of these materials were found. The results will help increase the amount of information about the properties of the materials under study and will allow the use of these compounds in a wider range of optoelectronic devices, in particular, in solar cells and other devices that use solar radiation to generate electric current.

**\*Corresponding author:** Dilshod Nematov, S.U. Umarov Physical-Technical Institute of the National Academy of Sciences of Tajikistan, Tajikistan. Email: dilnem@phti.ru

---

## 1. Introduction

One of the factors contributing to global warming is the burning of fossil fuels, which increases the concentration of carbon dioxide in the atmosphere, warming the planet and changing the climate (Whiteside & Herndon, 2023; Jones et al., 2023; Voumik et al., 2023a). The Earth absorbs a significant portion of the light emitted by the Sun when it reaches its surface. As a result of heating the planet with this energy, its surface emits infrared radiation. The greenhouse effect, which causes global warming, occurs when carbon dioxide in the atmosphere absorbs much of the incoming thermal radiation and reflects it back to the Earth's surface (Aakko-Saksa et al., 2023; Weart, 2023). It is known that electricity generation emits about 40% of greenhouse gases into the atmosphere (Andrews, 2023; Voumik et al., 2023b). However, the world now obtains 80% of its energy from fossil fuels (oil, coal, and gas), which also contribute to environmental pollution and the daily depletion of these natural supplies (Sinclair, 2020). Thus, in light of the anticipated need for electricity worldwide and the 21st - century green energy agenda, the issue of the development of alternative and renewable energy sources is brought up, and the potential for transforming non-traditional energy into electricity is examined (Kumar & Prajapati, 2023).

Alternative sources such as windmills, moisture-to-electricity converters, thermoelectric (TE) generators, photovoltaic (PV) converters (solar panels), solar thermophotovoltaic converters and the use of geothermal waters have a positive effect on the air environment, but they are not available everywhere. One of the effective natural sources of alternative energy is the synthesis and optimization of the properties of new nanomaterials to create solar energy converters into electricity, which is very effective and environmentally friendly compared to many other methods (Whiteside & Herndon, 2023; Jones et al., 2023; Voumik et al., 2023; Aakko-Saksa et al., 2023; Weart, 2023; Andrews, 2023; Voumik et al., 2023; Sinclair, 2020; Kumar & Prajapati, 2023).  The great growth potential of this alternative energy industry is due to such global factors as the need to ensure national energy security and the rising cost of fossil energy sources. Alternative energy has other unique advantages: for example, solar energy is available to everyone, free, practically inexhaustible, and the process of converting it into electrical energy does not have a negative impact on the environment (Zhang et al., 2023a; Yang et al., 2022; Fraser et al., 2023; Sayed et al., 2023).

In recent years, photovoltaics has been rapidly developing for the developed of which, in addition to silicon and perovskites, many types of crystalline materials have been developed and proposed (Luceño-Sánchez et al., 2019; Huang et al., 2021; Yao et al., 2020; Almora et al., 2023; Josephine et al., 2023; Bellucci et al., 2022; Rafin et al., 2023; Dada et al., 2023). One of these promising and promising materials is kesterite materials with the general formula $A_2BCD_4$ (A = Cu, Ag; B = Zn, Cd, Ni, Mg, ...; C = Sn, Ge, Si; D = S, Se) (Chen et al., 2010a). Since their emergence as so-called "Third generation generators," photovoltaic (PV) systems have been hailed as an environmentally and economically viable alternative to conventional technologies for solving the world's energy, safety and environmental problems (Chen et al., 2010b; Steinhagen et al., 2009; Chen et al., 2011). However, despite the obvious achievements in this area, the development and research of the fundamental properties of potentially new kesterite photovoltaic materials is of great importance for improving the performance of devices based on them.

Because of their direct bandgap in the 1.0–2.5 eV region, the most extensively used kesterite crystals based on the Cu2ZnSnS4 and Cu2ZnSnSe4 (CZTSSe) system are utilized in the industrial manufacture of solar panels (Persson, 2010). Kesterite's range of use is increased by the adjustable band gap, which also enables task-specific adjustment for ideal spectral matching (Haight et al., 2016). Furthermore, they are excellent candidates for the production of solar energy due to their p-type conductivity and high optical absorption coefficient ($>10^4$ cm$^{-1}$) (Fan et al., 2021). Additionally, this allows for the kesterite film to be thinned down to a point where the solar panel's cost can be decreased without sacrificing efficiency (Fan et al., 2021). The efficiency of kesterite-based solar cells has recently risen from 12.6% to 13.6% (Green et al., 2023; National Renewable Energy Laboratory, n.d.; Gong et al., 2022). However, the performance of solar cells based on this material is still far from the theoretical limit, indicating that the efficiency potential of kesterite is still little exploited. The reason for the poor performance of CZTSSe is mainly due to their high open circuit voltage deficit, which has been repeatedly reported in research papers (Gupta et al., 2019; Baid et al., 2021; Sahu et al., 2021; Pu et al., 2018). This is due to fluctuations in the band gap and potential induced by crystalline disorder between elements A and B sites of kesterite which occurs at the structural and electronic levels (Gupta et al., 2019; Baid et al., 2021; Sahu et al., 2021; Pu et al., 2018). Existing problems force researchers to develop new analogues of CZTSSe, however, obtaining new perovskites including some members of the Cu2NiXS4 family (X = Si, Ge, Sn) is labor-intensive work due to the complex single-phase growth of kesterite while obtaining a homogeneous and high-quality layer free of secondary phases (Sahu et al., 2021; Pu et al., 2018). Sometimes the final synthesis of kesterite materials results in many undesirable solid solutions, which complicate the work (Gao et al., 2014), and some resulting materials with a kesterite structure may have undesirable properties. Therefore, in recent years, preliminary prediction of properties applied to the synthesis of materials has become an integral tradition among the solid state community.

In this regard, recently the properties of kesterites have also been studied by various theoretical methods, as a result of which the efficiency of solar cells based on them is constantly increasing (Chen et al., 2011; Persson, 2010; Haight et al., 2016; Fan et al., 2021; Green et al., 2023;   National Renewable Energy Laboratory, n.d.; Gong et al., 2022; Gupta et al., 2019; Baid et al., 2021; Sahu et al., 2021; Pu et al., 2018; Gao et al., 2014). Density functional theory (DFT) is a potent theoretical approach that has gained significant traction in the last ten years as a major tool for the theoretical study of solid materials. Its potent approach accounts for the behavior of electrons in all atomic-molecular environments and offers a highly accurate reformulation of quantum mechanical calculations of solids. This is because contemporary computing clusters can solve the Kohn-Sham equations efficiently (Banerjee et al., 2015; Kato & Saito, 2023; Nematov, 2022; Nematov et al., 2022; Davlatshoevich, 2021; Zhang et al., 2023b; Zhang et al., 2023c; Annett, 1995; Kananenka, et al., 2013). However, these formulas are predicated on a single estimate, that of the exchange-correlation energy, which accounts for the precision of quantum computations. Here, many of the basic characteristics of compounds based on the Cu2NiXS4 (X = Si, Ge, Sn) family still remain poorly studied and correspond to the current topic requiring in-depth research, despite the annual increase in publications devoted to the study of the properties of kesterites.

In this work, using quantum chemical calculations within the framework of density functional theory, the electronic and optical properties of kesterites of the Cu2NiXS4 (X = Si, Ge, Sn) family are studied, a detailed study and disclosure of which is important for the appropriate selection of the synthesized material for specific applications.

## 2. Materials and Methods

Ab initio quantum chemical calculations within the framework of density functional theory were carried out on the basis of data on crystal lattices published in Nematov (2023), obtained after complete relaxation of lattice parameters using the VASP package (Kresse & Furthmüller, 1996) and the SCAN functional (Sun et al., 2015; Sahni et al., 1988; Perdew et al., 1996). In this case, electronic and ion relaxation was achieved at a cutoff energy of 550 eV and $3 \times 3 \times 3$ k-points. The band gap, electronic structure and optical properties of $Cu_2NiXS_4$ (X = Si, Ge, Sn) systems were studied using the VASP and Wien2k quantum chemical simulation codes using the hybrid functional HSE06 (Painter, 1981) and the modified TB-mBJ functional (Singh, 2010). The optimal plane wave cutoff value Kmax was selected as 6.0 $Ry^{1/2}$. The Kohn-Sham equations were solved using LAPW. Kesterite crystals of the tetragonal system (symmetry group I-4) were chosen as the structures under study. The following valence electrons were considered: Si: $3s^2\ 3p^2$, Cu: $3d^{10}\ 4s^1$, Sn: $4d^{10}\ 5s^2\ 5p^2$, Ni: $3d^{10}\ 4s^2$, Ge: $3d^{10}\ 4s^2\ 4p^2$, and S: $3s^2\ 3p^4$.

## 3. Results and Discussion

The energy band distribution diagram, the band energy's dependence on the density of electronic states (DOS), and the band gap's numerical values are used to evaluate the electronic characteristics of the $Cu_2NiXS_4$ (X = Si, Ge, and Sn) system. Table 1 compares the bandgap values we calculated within DFT - HSE06 with the results of experimental measurements by independent authors.

**Table 1**
**Comparison of the calculated value of the band gap of the $Cu_2NiSiS_4$, $Cu_2NiGeS_4$, $Cu_2NiSnS_4$ system with literature data**

| | Bandgap, eV | | |
|---|---|---|---|
| | **THIS WORK** | **LITERATURE** | |
| | **HSE06** | **Calc.** | **Experimental** |
| **$Cu_2NiSiS_4$** | 2.560 | - | - |
| **$Cu_2NiGeS_4$** | 1.802 | 1.13[a] | 1.8 [a] |
| **$Cu_2NiSnS_4$** | 1.321 | 1.26[b] | 1.31[c], 1.38[d] |

a.  (Beraich et al., 2020)
b.  (Chen & Persson, 2017)
c.  (Deepika & Meena, 2019)
d.  (Kamble et al., 2014)

As shown in Table 1, the results of bandgap calculations of $Cu_2NiSnS_4$, $Cu_2NiGeS_4$ crystals obtained from the HSE06 functional are in good agreement with experiment.

The total density of electronic states of $Cu_2NiSnS_4$, $Cu_2NiGeS_4$, and $Cu_2NiSiS_4$ semiconductor crystals were then analyzed using mBJ calculations. The results of calculations of the total density of electronic states for $Cu_2NiSnS_4$, $Cu_2NiGeS_4$, and $Cu_2NiSiS_4$ crystals are shown in Figure 1 (a-c).

**Figure 1**
**Total density of electronic states for Cu₂NiSiS₄ (a), Cu₂NiGeS₄ (b) and Cu₂NiSnS₄ (c)**

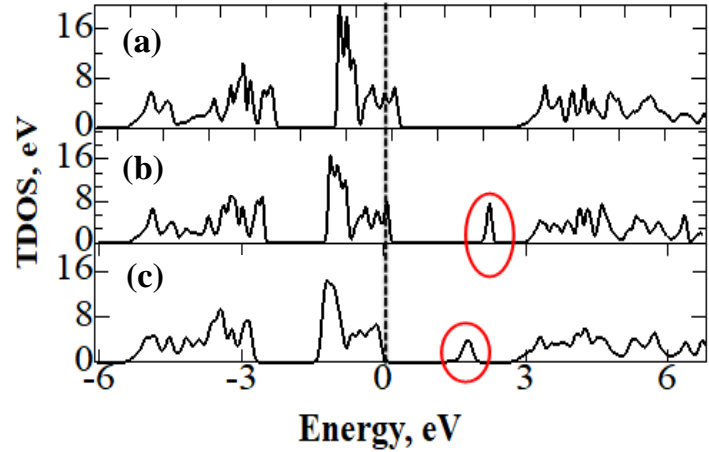

The results shown in Figure 2 demonstrate that the band gap lowers and the Fermi levels move towards the valence band when Si is substituted with Ge and Sn. Conversely, it is evident that the density of states drops when Sn takes the place of Si. In this case, new electronic states are formed in the energy gap of Cu₂NiGeS₄ and Cu₂NiSnS₄, which are important from the point of view of using the material in electronic devices.

To justify the change in the bandgap width, it is necessary to analyze the spectra of the optical properties of the materials under study. Calculated optical properties of materials, including their absorption coefficient and refractive index, provide information about what type of response these materials will exhibit when photons are incident on them (Qiu et al., 2017). The optical properties of the Cu₂NiXS₄ (X = Si, Ge, Sn) system were investigated based on the calculation of their real ($\varepsilon 1$) and imaginary ($\varepsilon 2$) parts of the dielectric functions. The real part shows the energy-saving capacity of a material, which is something that is assumed to be inherent in all materials at zero energy or zero frequency limit. Figure 2 (a, b) shows the curves of $\varepsilon 1$ and $\varepsilon 2$ versus the energy of incident photons for kesterites of the Cu₂NiXS₄ (X = Si, Ge, Sn) family.

**Figure 2**
**Real (a) and imaginary part (b) dielectric constant of kesterites**
**of the Cu₂NiXS₄ (X = Si, Ge, Sn) family**

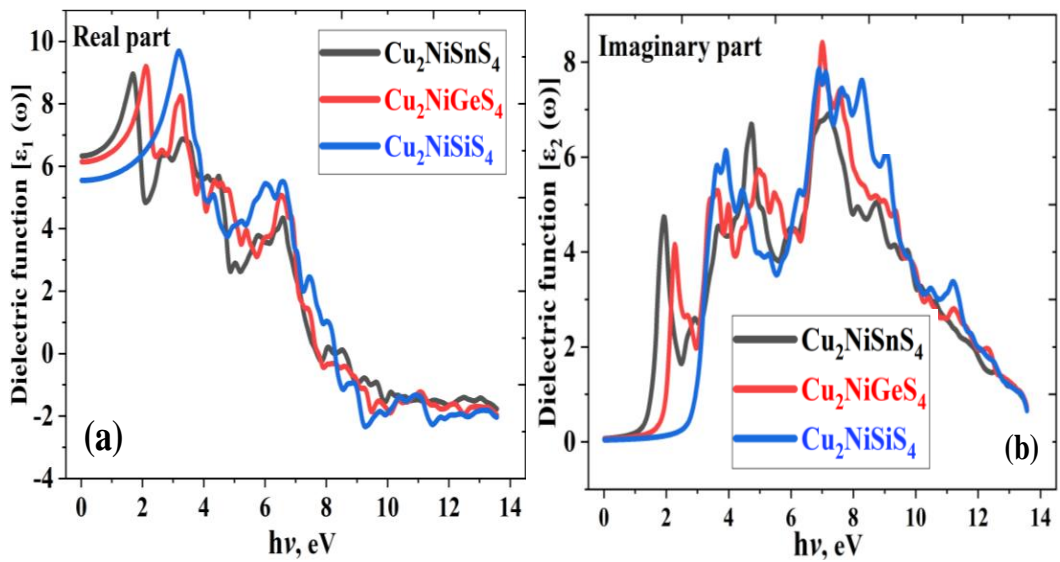

From Figure 2(a) it can be seen that at the highest photon energies, all these materials, namely kesterite containing silicon, exhibit metallic behavior. That is, a negative value of the real part indicates the possession of a metallic nature at high energies. This makes it possible to estimate the metallicity fractions of materials using a real function, which shows feedback from the optical band gap. For solar devices, the behavior of these materials indicates the energy gap, whereas the imaginary component of the dielectric function indicates the compound's absorptive capacity. This provides information about how the material reacts when exposed to electromagnetic radiation (Zafar et al., 2019; Johnson & Guy, 1972; Jia et al., 2023; Qian et al., 2023). According to Figure 2(b), the replacement of Si with Ge and Sn leads to an increase in the absorption coefficient of the materials under study in the IR and visible radiation range, which is important for the use of materials in solar panels.

**Figure 3**
**Calculated refractive index spectra of the $Cu_2NiXS_4$ (X = Si, Ge, Sn) system as a function of photon energy**

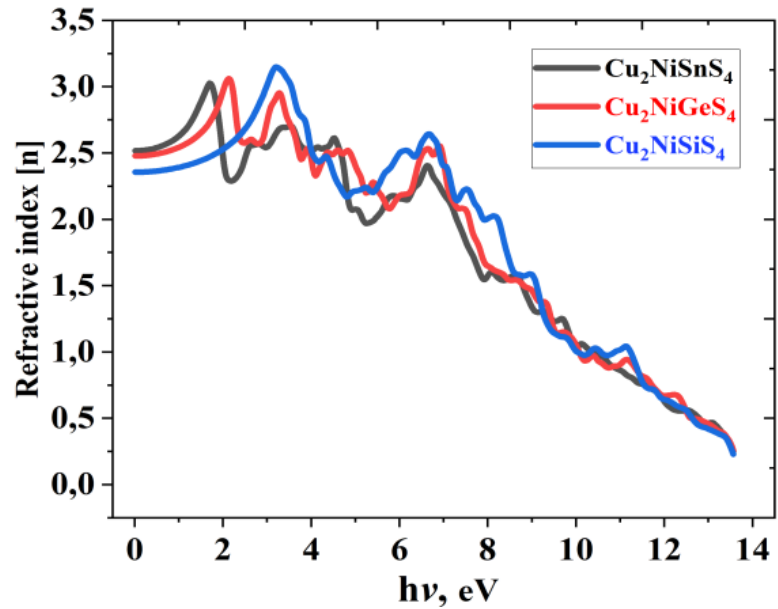

The results displayed in Figure 3 make it evident that adding germanium and tin in place of silicon raises the $Cu_2NiXS_4$ (X = Si, Ge, Sn) system's refractive index (n). In some energy ranges, the refractive index falls drastically below unity after reaching its maximum value. Furthermore, we can deduce from the expression n=c/9 that a refractive index value less than one means that the incident radiation's phase velocity is greater than c, which allows the incident rays to pass through the material and turn it transparent to incoming radiation (Kortüm, 2012; Modest & Mazumder, 2021). Table 2 shows the static values of ε1x(0), ε2z(0) and n according to the DFT-mBJ-WIEN2k calculations.

**Table 2**
**Calculated values of statistical dielectric constant and refractive index for kesterites of the $Cu_2NiXS_4$ (X = Si, Ge, Sn) family**

| System | $\varepsilon_1^x(0)$ | $\varepsilon_2^z(0)$ | $n$ |
|---|---|---|---|
| $Cu_2NiSiS_4$ | 5.68 | 5.61 | 2.52 |
| $Cu_2NiGeS_4$ | 6.11 | 6.46 | 2.48 |
| $Cu_2NiXSnS_4$ | 6.46 | 6.88 | 2.36 |

Beyond the Fermi level, it is known that photon absorption excites occupied states toward unoccupied states. It is known as "optical conduction" when the photons cross the interband transition and as "interband absorption" when they are absorbed. When light is subjected to an electric field, conductivity is known as optical conductivity. Figures 4 and 5 display the computed spectra of the real and imaginary components of the optical conductivity of the systems that are being studied. It is also evident from these spectra that adding Ge and Sn to Si results in improved photoconductivity.

**Figure 4**
**Photon energy-dependent spectra of the real part of the optical conductivity of the Cu₂NiXS₄ (X = Si, Ge, Sn) system**

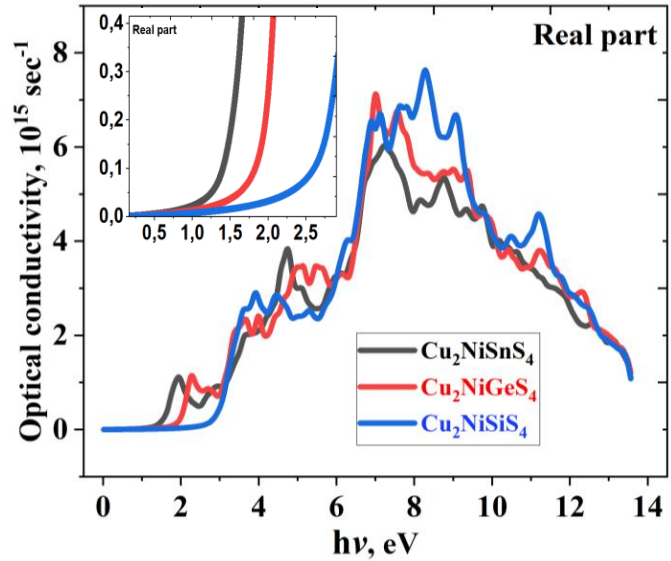

**Figure 5**
**Energy-dependent spectra of the imaginary part of the optical conductivity of photons of the Cu₂NiXS₄ (X = Si, Ge, Sn) system**

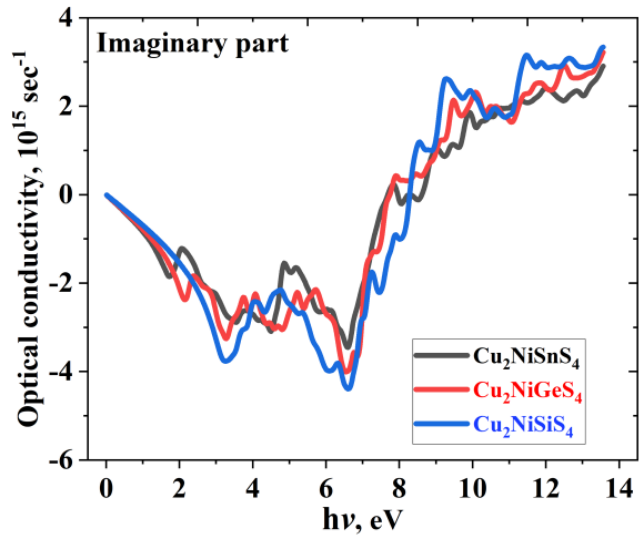

Judging by Figures 4 and 5, optical conductivity for the Cu₂NiGeS₄ and Cu₂NiSnS₄ system begins at energies lower than in the case of Cu₂NiSiS₄. From the data shown in Figure 5, it is clear thet al. these materials actively absorb light even at low photon energies, namely Cu₂NiSnS₄ is sensitive even to rays with an energy of 0.7 eV. Cu₂NiSiS₄ also begins to be activated in the energy range above 1.3 eV, having the highest photoconductivity when absorbing short-wavelength radiation (Schäfer & Nitsche, 1974;

Mani et al., 2023; Khatun et al., 2023; Kolhe et al., 2023; Davlatshoevich et al., 2023; Khouja et al., 2023; Nizomov et al., 2023; Nematov et al., 2023; Nematov et al., 2024; Nematov et al., 2021).

## 4. Conclusion

The mBJ exchange correlation potential is employed in this study to simulate the electrical and optical characteristics of $Cu_2NiXS_4$ (X = Si, Ge, Sn). Even though the $Cu_2NiXS_4$ (X = Si, Ge, and Sn) system's members differ in composition and structure, the permittivity curves and primary optical spectra of all of them exhibit very comparable properties in the infrared and visible radiation ranges. The optical absorption coefficient (>$10^4$ cm$^{-1}$) in the infrared and visible light energy ranges has been discovered to be relatively substantial, and it is proportional to the imaginary part of the permittivity.

## Funding Support

The work was supported financially by the International Foundation for Humanitarian Cooperation of the CIS within the framework of a scientific project at the expense of a grant from the International Innovation Center for Nanotechnologies of the CIS (GRANT No. 23-112).

## Ethical Statement

This study does not contain any studies with human or animal subjects performed by any of the authors.

## Conflicts of Interest

The author declares that he has no conflicts of interest to this work.

## Data Availability Statement

Data sharing is not applicable to this article as no new data were created or analyzed in this study.